\documentclass[final,5p,times,twocolumn]{elsarticle}

\usepackage{graphicx}%
\usepackage{multirow}%
\usepackage{amsmath,amssymb,amsfonts}%
\usepackage{amsthm}%
\usepackage{mathrsfs}%
\usepackage[title]{appendix}%
\usepackage{xcolor}%
\usepackage{textcomp}%
\usepackage{manyfoot}%
\usepackage{booktabs}%
\usepackage{algorithm}%
\usepackage{algorithmicx}%
\usepackage{algpseudocode}%
\usepackage{listings}%
\usepackage[breaklinks=true,colorlinks,citecolor=blue,linkcolor=blue,urlcolor=blue]{hyperref}

\begin{document}
\begin{frontmatter}

\title{Microstructural Studies Using Generative Adversarial Network (GAN): a Case Study}

\author[a]{Owais Ahmad}
\author[a]{Vishal Panwar}
\author[b]{Kaushik Das}
\author[a]{Rajdip Mukherjee}
\author[a]{Somnath Bhowmick*\corref{cor}}
\ead{bsomnath@iitk.ac.in}

\address[a]{Department of Materials Science and Engineering, Indian Institute of
Technology, Kanpur, Kanpur-208016, UP, India}
\address[b]{Department of Metallurgy and Materials Engineering, Indian Institute of Engineering Science and Technology, West Bengal, Shibpur, Howrah, 711103, India}

\begin{abstract}
The generative adversarial network (GAN) is one of the most widely used deep generative models for synthesizing high-quality images with the same statistics as the training set. Finite element method (FEM) based property prediction often relies on synthetically generated microstructures. The phase-field model is a computational method of generating realistic microstructures considering the underlying thermodynamics and kinetics of the material. Due to the expensive nature of the simulations, it is not always feasible to use phase-field for synthetic microstructure generation. In this work, we train a GAN with microstructures generated from the phase-field simulations. 
Mechanical properties calculated using the finite element method on synthetic and actual phase field microstructures show excellent agreement. Since the GAN model generates thousands of images within seconds, it has the potential to improve the quality of synthetic microstructures needed for FEM calculations or any other applications requiring a large number of realistic synthetic images at minimal computational cost.  
\end{abstract}


\end{frontmatter}
\section{Introduction}
\label{intro}
The properties and performance of a material are directly linked to its microstructure. The finite element method (FEM) is a popular technique for microstructure-based property prediction. Prominent examples are crystal plasticity finite-element modeling~\cite{ROTERS20101152,abaqus}, image-based finite-element analysis~\cite{Langer200115}, and combined phase-field and finite-element modeling for structure–property correlation~\cite{FROMM20125984, Linda_2024}. It is a common practice to generate synthetic microstructures for FEM analysis~\cite{LIU2020102614, dream}. Synthetic microstructures are often used, as getting a large number of experimental microstructures is an expensive and time-consuming process.

Besides experimental methods like optical, scanning, and tunneling electron microscopy, computational methods like phase-field modeling (PFM) are widely used to study the microstructure of a material. PFM is a versatile computational tool that has been used to study the microstructure evolution during precipitate growth~\cite{mukherjee2009, Mukhrjee2010, Gururajan2007} and grain growth~\cite{CHANG201767,verma_mukherjee, PhysRevLett.86.842}, coarsening~\cite{MOLNAR20126961}, solidification~\cite{ZHAO20191044, chatterjee2008phase}, spinodal decomposition~\cite{CHAFLE}, and phase-separation~\cite{RAGHAVAN2021}. Implementing PFM is computationally intensive because it involves solving partial differential equations numerically. Attempts have been made to reduce computational time by employing advanced algorithms (e.g., active parameter tracking), and high-performance computational architecture (e.g., parallel computing and GPU-accelerated PFM) ~\cite{vondrous2014parallel}.

Artificial intelligence and machine learning (ML) have been used in various fields of materials science and engineering~\cite{wang2020}, including phase-field simulations and microstructure modeling~\cite{xue2022physics, oommen2022learning, WU2023}. Recently, several methods have been proposed for accelerating the prediction of microstructure evolution via machine-learned surrogate models~\cite{hu2022accelerating, montes2021accelerating, TIWARI2025113518}. These models are based on recurrent neural networks (RNNs) and long short-term memory (LSTM) networks, trained to predict the future based on the output from the previous steps. Even after complete training, a model requires some microstructures from the actual phase-field simulation to predict new microstructures. Moreover, RNN and LSTM have limitations regarding how far the model can predict accurately based on past data. As a result, to study microstructure evolution for a very long time, one needs to switch back to the phase-field model at specific intervals to maintain high accuracy~\cite{ahmad2023accelerating, ahmad2024}. 

\begin{figure*}[t]
\centerline{\includegraphics[width=\textwidth]{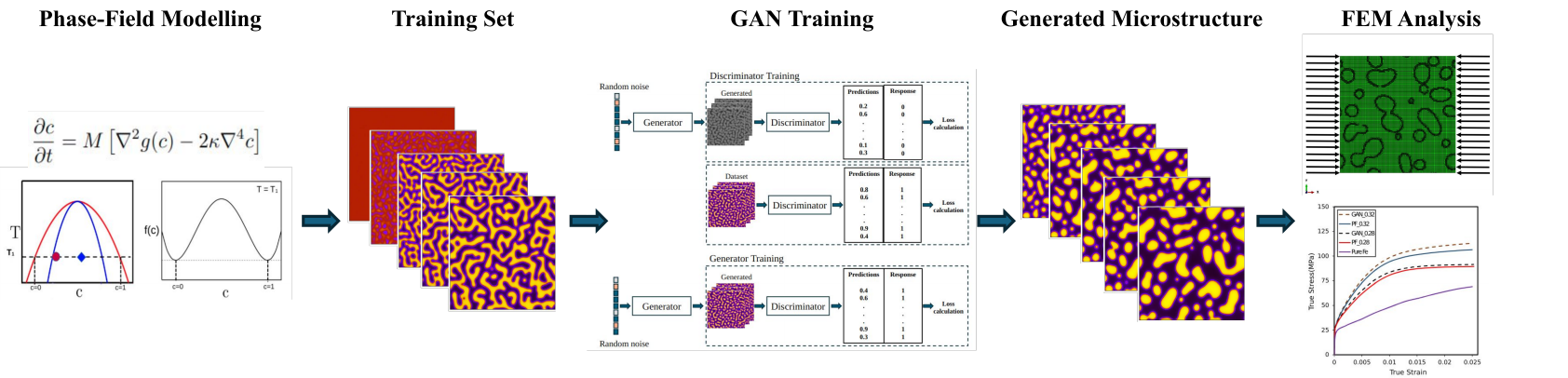}}
\caption{Schematic diagram of accelerated microstructure-property correlation: training data generation, GAN model training, synthetic microstructure prediction, and FEM analysis.}
\label{fig0}
\end{figure*}

In this work, we propose a model that is not hindered by the above-mentioned limitations. The model is based on a generative adversarial network (GAN), which is trained using phase-field generated microstructures for a wide range of compositions and time steps. Once trained, the model can generate realistic microstructures, starting with random noise as input. Thus, unlike the RNN or LSTM-based studies, the trained GAN model does not require phase-field-generated images to predict new microstructures. 
The deep generative model can generate thousands of images within a few seconds, proving it a viable option for creating an extensive database of synthetic, realistic microstructures.

\begin{figure*}
\centerline{\includegraphics[width=\textwidth]{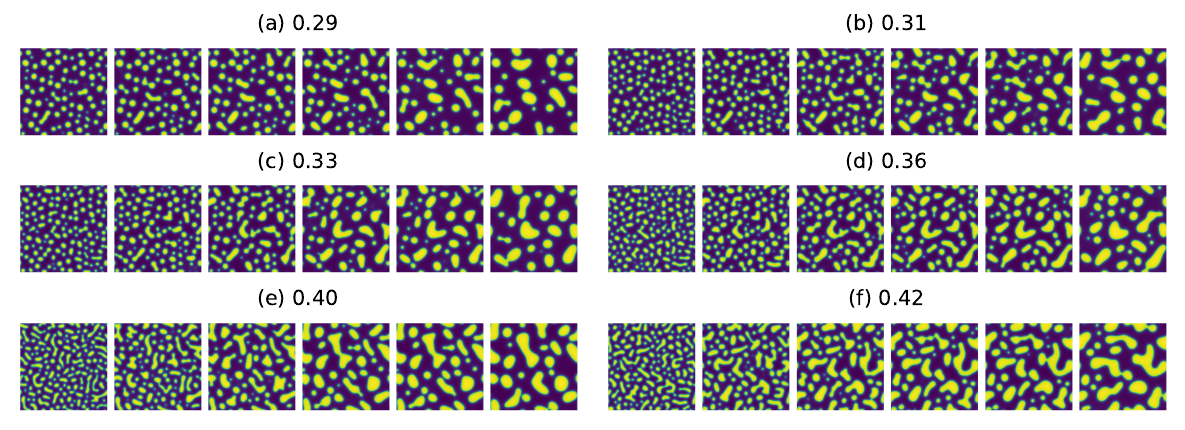}}
\caption{Microstructures for different phase fractions captured during the spinodal decomposition. Time evolution is simulated by the phase-field method.}
\label{fig1}
\end{figure*}

We also demonstrate the utility of synthetic microstructures for the structure-property correlation of materials. We use FEM to calculate the elastoplastic properties using the synthetic microstructures generated by GAN, and they are in excellent agreement with the results obtained using the actual phase-field microstructures. The ability of GAN to generate diverse and realistic microstructures can significantly accelerate material discovery and development. 

Recent advancements in deep learning have led to the development of sophisticated architectures that go beyond conventional GANs. Models incorporating dynamic collaborative adversarial domain adaptation networks, residual-self-attention feature fusion networks, and residual frequency attention regularized frameworks have demonstrated enhanced capability in learning complex representations and improving the fidelity of generated outputs~\cite{westphal2024generative, LIU2023798, Vinodhini_2025}. While this work primarily employs a standard GAN architecture, the integration of such advanced mechanisms presents promising directions for future research, although they come with a high computational cost. These developments reflect the broader trajectory of deep generative modeling in materials science, aiming for higher accuracy, adaptability, and computational efficiency.

Figure~\ref{fig0} schematically illustrates the workflow. The training data set is generated using a phase-field model for a chosen alloy. One needs to train two competing neural network models, a generator and a discriminator. The trained model generates thousands of microstructures, spanning a wide composition range and time, for the chosen microstructure. The computational cost for generating such a large number of images using GAN is negligible compared to the actual phase-field simulations. GAN-generated synthetic microstructures are finally subjected to FEM analysis for property prediction. In the following text, we first describe the technical details of PFM, GAN, and FEM, followed by a section on results and discussions, and conclusions.

\section{Methodology}
\label{method}
\subsection{Training dataset: Phase-field simulation}
The current study utilizes a phase-field model to create a training dataset of microstructure evolution during spinodal decomposition in an $A-B$ binary alloy. The total free energy of the alloy is ~\cite{doi:10.1063/1.1744102}, 
\begin{equation}
F=\int_V[f(c)+\kappa(\nabla c)^2] dV.
\label{tot_free_energy}
\end{equation}
The conserved phase-field variable $c(\boldsymbol{r},t)$, where $\boldsymbol{r}$ and $t$ denote space and time, represents the local composition of the system. A double-well potential describes the bulk free energy density, 
\begin{equation}
f(c)=W c^2 (1-c)^2.
\label{free_energy_density}
\end{equation}
The constant $W$ determines the potential barrier height between the two equilibrium phases corresponding to the compositions  $c=0$ and $c=1$, respectively. The total free energy also contains the gradients in the local composition term, given by  $\kappa (\nabla c)^2$ in Equation~\ref{tot_free_energy}, where $\kappa$ is the gradient energy coefficient.  Spinodal decomposition occurs when the composition lies between the two inflection points where $\partial^2f/\partial c^2<0$.  The Cahn-Hilliard equation~\cite{cahn1961spinodal} governs the spatiotemporal evolution of the conserved phase-field variable $c(\boldsymbol{r},t)$, 
\begin{align}
\dfrac{\partial c}{\partial t}=M \left[\nabla^2 g(c)-2\kappa\nabla^4 c\right].
\label{CH}
\end{align}
In the above equation, $M$ is the constant atomic mobility, and $g(c)=\partial f/\partial c$. Section S1 of the Supporting Information (SI) describes other details about implementing the phase-field model. 

In this study, we focus on the spinodal decomposition of a binary system. The phase fraction value $\phi_A$ ranges from 0.28 to 0.43, and  $\phi_B=1-\phi_A$. The phase mobilities for both components are set to 1. We perform the simulations on a 2D square domain, discretized with $256 \times 256$ grid points, and we allow the microstructure to evolve for more than 1000 time steps. Figure~\ref{fig1} shows selected phase-field microstructures with different initial compositions. We store the microstructures every single time step for each composition. The dataset contains approximately 20000 microstructure evolution images of $256 \times 256$ resolution. Since the simulations start with an initial composition described by random noise, the microstructures have no recognizable features from frame $t_{0}$ to $t_{20}$, which are not included in the dataset used for training the GAN.

\begin{figure*}[t]
\centering
\includegraphics[width=0.8\textwidth]{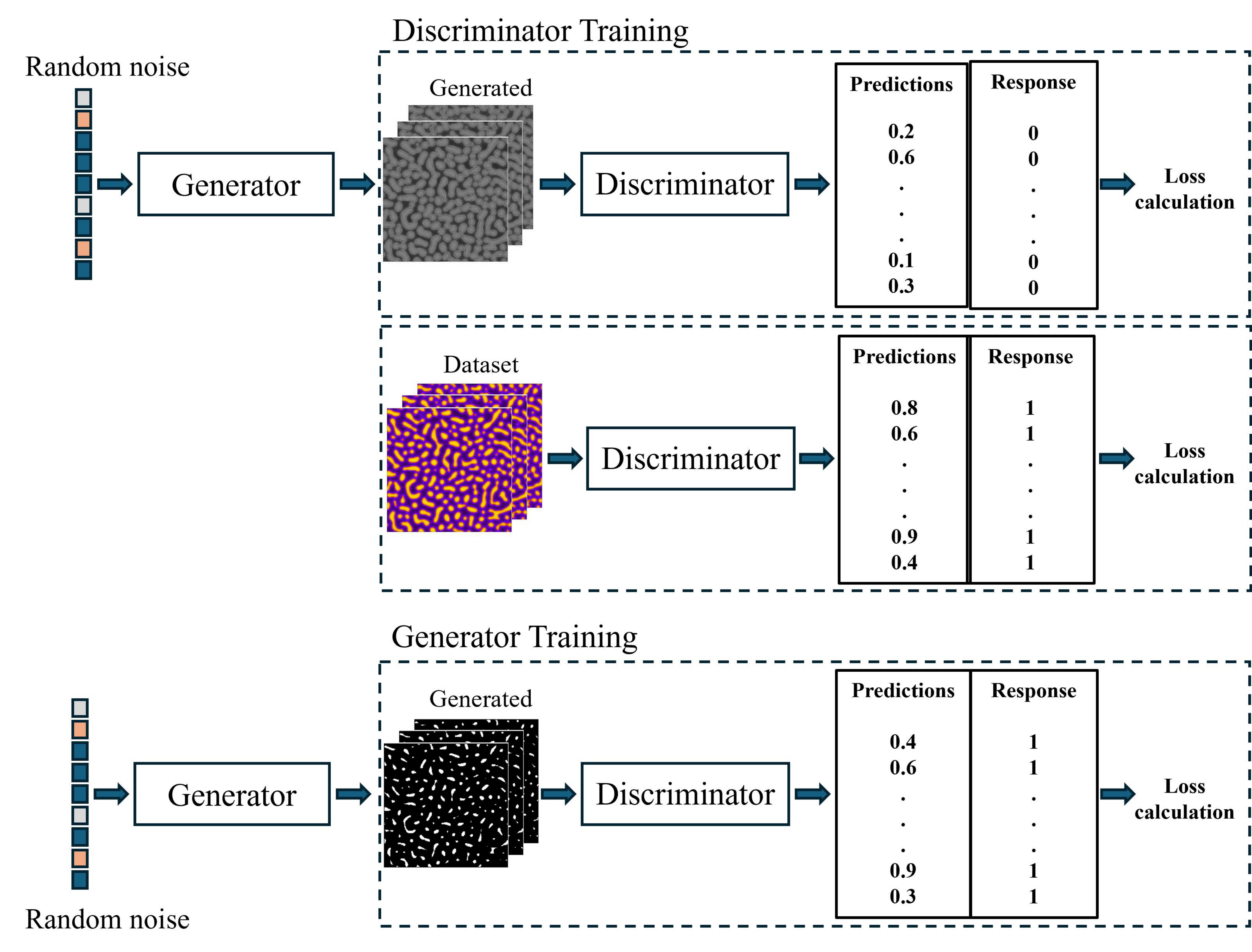}
\caption{Schematic representation of a Generative Adversarial Network (GAN). The diagram illustrates the adversarial training process where the generator synthesizes data, attempting to mimic a target dataset, depicted in a sequence of transformations from left to right. Concurrently, the discriminator evaluates the data, distinguishing between authentic and generated samples. This iterative process, shown through sequential stages, helps refine the generator's output to produce increasingly realistic data, as guided by the feedback from the discriminator.}
\label{fig2}
\end{figure*}

\subsection{Model training: Generative adversarial networks}
The usefulness of generative adversarial networks (GANs) in data science is profound. GAN has been successfully applied in various domains like image generation, video creation, and text generation, producing results that are often indistinguishable from the actual data. In the context of microstructure generation, GAN has significant potential. GAN can learn the underlying distribution of microstructural patterns and generate new synthetic microstructures statistically similar to the actual microstructures~\cite{narikawa2022generative, HENKES2022115497}. This capability is invaluable for materials scientists, as synthetic microstructures can predict material properties, optimize manufacturing processes, or explore new material designs that might not be easily accessible through traditional experimental or simulation methods. To the best of our knowledge, the capability of GAN to predict phase-field-like microstructure evolution remains unexplored. The rest of the section details the step-by-step implementation of GAN to capture the time evolution of spinodal decomposition for a wide range of compositions.

Generative Adversarial Networks (GAN) operate on a unique principle that involves two competing neural network models: a generator (G) and a discriminator (D). The generator aims to produce data that mimics a target distribution, while the discriminator evaluates the authenticity of the data, differentiating between actual and generated samples. The schematic in Figure~\ref{fig2} illustrates the GAN operation through a series of data transformations and evaluation steps. GAN embodies a dual-component framework comprising a generator and a discriminator, which are trained concurrently through an adversarial process. The generator synthesizes data that approximates authentic datasets' characteristics, as visualized in the sequential images from left to right in Figure~\ref{fig2}. Simultaneously, the discriminator assesses whether the incoming data resembles genuine or synthesized datasets, providing feedback that guides the iterative refinement of the generator.

During the training phase, as depicted in the discrete stages of Figure~\ref{fig2}, the generator produces data instances that the discriminator evaluates. The initial outputs from the generator are often rudimentary and easily distinguishable from the target dataset. However, as training progresses, the discriminator's ability to classify data as accurate or generated imposes a sophisticated learning challenge, compelling the generator to produce increasingly realistic data. This adversarial training continues until a state of equilibrium is reached where the discriminator's ability to differentiate real from generated data is akin to random guessing, indicating that the generated data are nearly indistinguishable from the actual data. This process is illustrated in Figure~\ref{fig2}, where each panel represents a snapshot in the evolutionary timeline of data generation, highlighting the gradual enhancement in data quality as the generator adapts to the discriminator's feedback. The training process involves optimizing both networks. The generator, \( G \), takes a random noise vector, \( z \), as input and generates a data sample, \( G(z) \). The discriminator, \( D \), takes a data sample (either real or generated) as input and outputs a probability score representing the likelihood of the sample being real. The objective function for a GAN is formulated as:
\begin{equation}
\begin{aligned}
\underset{G}{\min} \, \underset{D}{\max} \, V(D, G) = & \mathbb{E}_{x \sim p_{\text{data}}(x)}[\log D(x)] \\ & + \mathbb{E}_{z \sim p_{z}(z)}[\log(1 - D(G(z)))].
\end{aligned}
\end{equation}
Here, \( \mathbb{E}_{x \sim p_{\text{data}}(x)}[\log D(x)] \) is the expected value of the logarithm of the discriminator's output for accurate data \( x \), and \( \mathbb{E}_{z \sim p_{z}(z)}[\log(1 - D(G(z)))] \) is the expected value of the logarithm of one minus the discriminator's output for generated data. The discriminator aims to maximize this function by better distinguishing actual data from fake data better. In contrast, the generator tries to minimize this function, aiming to fool the discriminator into thinking its generated data is real. During training, the discriminator's parameters are adjusted to maximize \( V(D, G) \), whereas the generator's parameters are adjusted to minimize \( V(D, G) \). This training process leads to a scenario where the generator learns to produce increasingly realistic data, and the discriminator becomes better at distinguishing actual data from these high-quality fake samples. Ideally, the equilibrium of this training process is when the generator produces data indistinguishable from actual data, and the discriminator is at a 50 \% accuracy level, unable to differentiate between actual and generated data. 

To ensure proper GAN convergence, the generator and discriminator losses are tracked to reach a relative equilibrium, where neither loss consistently dominates the other, and both oscillate around stable points. Additionally, we generate and visually inspect a set of synthetic microstructures every 100 epochs to evaluate the quality and diversity of generated samples, ensuring consistent feature reproduction and the absence of mode collapse. This dual monitoring approach of loss stability and regular visual inspection helps confirm the model's convergence and reliability in generating realistic microstructures.

\subsection{Architecture Description with Wasserstein Loss}

\begin{table}[h]
\small
\renewcommand{\arraystretch}{1.3}  
\centering
\caption{Generator and Discriminator Architecture with Hyperparameters}
\vspace{4pt}
\begin{minipage}{\columnwidth}  
\centering
\setlength{\tabcolsep}{4pt}  
\begin{tabular}{|l|l|l|}
\hline
\textbf{Model} & \textbf{Layer} & \textbf{Details} \\
\hline
\multirow{5}{*}{\parbox{1cm}{Gen}} & Dense + BN + LReLU & 64×64×256 units \\
& Reshape & (64, 64, 256) \\
& ConvT + BN + LReLU & (4×4), s=2, 128 filters \\
& ConvT + BN + LReLU & (4×4), s=2, 64 filters \\
& ConvT + tanh & (4×4), s=1, 1 filter \\
\hline
\multirow{5}{*}{\parbox{1cm}{Disc}} & Conv + LReLU + Drop & (5×5), s=2, 32 filters, d=0.3 \\
& Conv + LReLU + Drop & (5×5), s=2, 64 filters, d=0.3 \\
& Conv + LReLU + Drop & (5×5), s=2, 128 filters, d=0.3 \\
& Flatten & \\
& Dense & 1 unit (binary) \\
\hline
\multicolumn{3}{|l|}{\textbf{Hyperparameters}} \\
\hline
\multicolumn{2}{|l|}{Loss Function} & Binary Cross-Entropy \\
\multicolumn{2}{|l|}{Optimizers} & Adam (LR=2e-4) \\
\hline
\end{tabular}
\end{minipage}
\vspace{2mm}  
\end{table}

The presented architecture implements a Wasserstein GAN (WGAN) for generating grayscale images of size 256×256. The WGAN utilizes the Wasserstein distance (also known as Earth Mover's distance) instead of the traditional Jensen-Shannon divergence used in standard GANs, which helps provide more stable training and mitigates mode collapse.

The generator network employs a series of transposed convolutions to progressively upsample the input from a dense layer to the target image dimensions. Each upsampling stage is followed by batch normalization and LeakyReLU activation, except for the final layer, which uses tanh activation to produce normalized pixel values. The architecture enables smooth upsampling from a 64×64 feature map to the final 256×256 image through two stride-2 transposed convolutions.

The discriminator (critic) follows a conventional convolutional architecture with three convolutional layers, each reducing spatial dimensions while increasing feature channels. Notably, the discriminator implements several key WGAN principles, like: (a) no sigmoid activation in the final layer, allowing unbounded critic values, (b) Dropout layers (0.3) after each convolution to enhance regularization, and (c) LeakyReLU activations to prevent sparse gradients. The Wasserstein loss is computed as:
\begin{equation}
    L = \mathbb{E}_{x \sim \mathbb{P}_r}[D(x)] - \mathbb{E}_{z \sim \mathbb{P}_z}[D(G(z))]
\end{equation}
where $D$ represents the critic (discriminator), $G$ the generator, $\mathbb{P}_r$ the real data distribution, and $\mathbb{P}_z$ the noise distribution. To ensure the Lipschitz constraint required by the Wasserstein metric, gradient clipping or gradient penalty should be implemented during training. The training is conducted on an Intel\textregistered{} Xeon\textregistered{} Gold 6338N CPU (2.20~GHz).

This approach can be extended to 3D microstructures; however, it is heavily dependent on the availability of computational resources. The extension would begin with a 3D noise input vector for the generator, which would be adapted to predict 3D microstructures. The discriminator would then process 3D ground truth data to evaluate the generated 3D outputs. In the current work, the focus has been on 2D due to computational cost limitations.

We would like to point out that GANs are well-suited for generating sharp and realistic microstructures due to their adversarial training framework, which promotes high-fidelity outputs. In contrast, while diffusion models offer better mode coverage and training stability~\cite{azqadan2023predictive}, they generally require longer training and inference times than GANs.

\subsection{Virtual Mechanical Testing}
\begin{figure*}[t]
\centering
\includegraphics[width=0.9\textwidth]{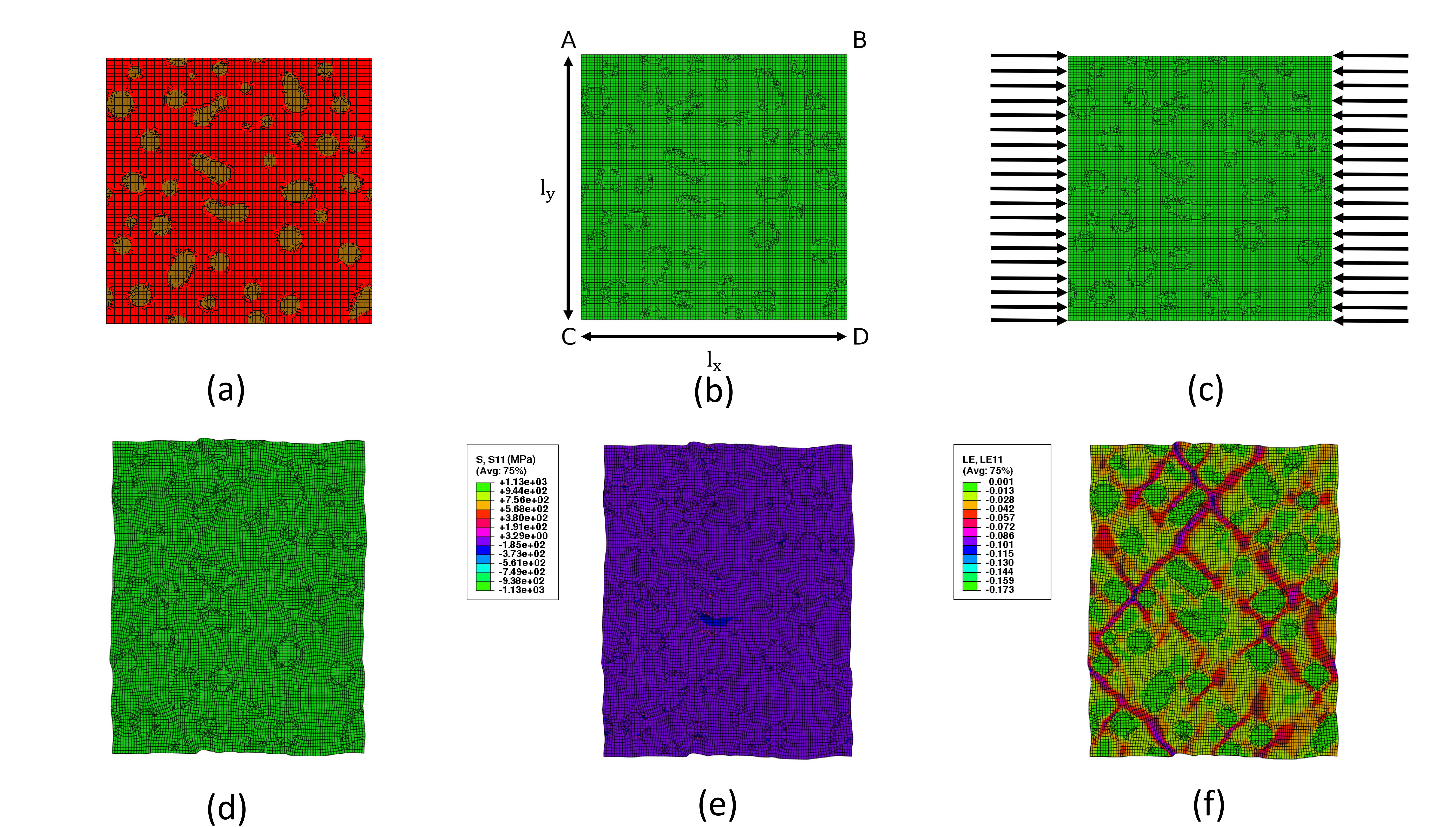}
\caption{Methodology for computing elastoplastic behavior: (a) Meshed microstructure from OOF2, (b) Categorization of nodes on the imported microstructure in Abaqus for applying periodic boundary condition, (c) Schematic for applying unidirectional uniform compressive strain in x-direction on imported microstructure, (d) Deformed microstructure at the scale of 5X deformation factor, (e) Distribution of true stress $\sigma_{xx}$ in the deformed microstructure, and (f) distribution of true strain $\epsilon_{xx}$ for deformed microstructure. The stresses and strain shown here are evaluated at the final time step.}
\label{methodology}
\end{figure*}
A simple homogenization approach, commonly used for composites, is adopted to compare the mechanical behavior under compression for microstructures from GAN and phase-field methods under similar conditions. In this approach, the area-averaged stress component for the entire microstructure is evaluated for every value of the applied strain component. For this purpose, the matrix is assumed to be pure iron, and the precipitate is assumed to be pure chromium. The mechanical properties for pure iron and chromium single crystal at room temperature have been taken from literature~\cite{Fritz2017176, Yabe20141342, hertzberg2012deformation} and are given in Section S2, Table S1 and Table S2 of the Supporting Information (SI). 

The weak form of the force equilibrium equation $\nabla\sigma = 0$, subject to periodic boundary conditions, is solved using the finite element method. The mesh for finite element analysis (FEA) of microstructure is directly prepared from the image of microstructure with OOF2~\cite{Langer200115}, an open-source software package. First, we identify the phases present in the microstructure. An adaptive periodic mesh, according to the morphology of the microstructure, is formed from the initial skeleton of the microstructure with the help of the OOF2 skeleton modification tool, as shown in Figure~\ref{methodology}(a). A mesh convergence study was performed, and the optimum initial skeleton size was found to be $100 \times 100$ rectangular finite elements grid. The details of the convergence study are given in Section~S2-B and Figure~S1 of the Supporting Information. 

Then, the meshed microstructure is imported in Abaqus~\cite{abaqus}, a commercial finite element analysis software, for further elastoplastic analysis. 
In Abaqus, the elastic and the plastic properties are assigned to individual phases, i.e., pure iron to matrix and pure chromium to precipitate. We apply periodic boundary conditions on the imported microstructure mesh using boundary nodes through a linear constraint equation,
\begin{equation}
A_{m}u^p_{i}+A_{n}u^q_{j}+\cdots+A_{o}u^r_{k} = 0,
\label{PBE}
\end{equation}
where, $A_m,A_n...$ are the coefficients, $u$ is the nodal variable-displacement, $i,j...$ is the degree of freedom, and $p,q...$ are the node numbers. Further details regarding the choice of boundary condition are given in Section~S2-C and Figure~S2 of the Supporting Information. The constraint equations are applied to mesh through reference points (RP), and hence the constraint equations become:
\begin{equation}
A_{m}u^p_{i}+A_{n}u^q_{j}+\cdots+A_{o}u^r_{k} = u^{RP}_{i}.
\label{PBE_M}
\end{equation}
The boundary nodes of meshed microstructure are categorized to apply the constraint equations. Two different sets of nodes are created: one as corner nodes - A, B, C, and D, and the other as edge nodes AB, AC, BD, and CD, as shown in Figure~\ref{methodology}(b). The size of imported microstructure is $l_x\times l_y$, where $l_{x} = l_{y} = 256$ nm. The three reference points RP1, RP2, and RP3 are created outside of the meshed microstructure and act as dummy nodes. The constraint equations for different nodal sets are given in Table \ref{corner} and Table \ref{edge}. 

The microstructure is tested with plain strain assumption to simulate the elastoplastic response. The CPE3 and CPE4 element types are used for triangular and quadrilateral elements in plain strain conditions. A static step is created with an initial increment of $10^{-4}$. The minimum and maximum increment sizes of $10^{-9}$ and  $10^{-1}$ are assigned to the static step, respectively, with a large number of iterations, in the order of $10^{6}$ to prevent any solver error due to insufficient number of iteration. Chromium is a brittle material at room temperature. Hence, we choose to test the microstructure in compression with small compressive strains without considering any damage mechanism. We apply a uniform compressive strain of 0.01 through reference point RP1 as given in Table \ref{BC} on the AC and BD edges of the microstructure in the x direction while the center of the microstructure is fixed. The top and bottom edges are free to deform. Figure~\ref{methodology}(c) shows a schematic of boundary conditions. The Abaqus~\cite{abaqus} implicit solver is used to solve the resulting finite element formulation. Figure~\ref{methodology}(d) shows the deformed microstructure with a 5X deformation scale factor. 
\begin{table}[ht]
\caption{Constraint equation on corner nodes}
\centering
\begin{tabular}{c@{\hskip 0.5in}c} 

\hline
\textbf{B and A} & \textbf{D and C} \\ 
\hline
$u_{x}^{B} - u_{x}^{A} -l_{x}u_{x}^{RP1}=0$& $u_{x}^{D} - u_{x}^{C} -l_{x}u_{x}^{RP1}=0$\\

$u_{y}^{B} - u_{y}^{A} -\frac{l_{x}}{2}u_{x}^{RP3}=0$&$u_{y}^{D} - u_{y}^{C} -\frac{l_{x}}{2}u_{x}^{RP3}=0$     \\
\hline
\multicolumn{2}{c}{\textbf{A and C}} \\
\hline
\multicolumn{2}{c}{$u_{x}^{A} - u_{x}^{C} -\frac{l_{y}}{2}u_{x}^{RP3}=0$}   \\

\multicolumn{2}{c}{$u_{y}^{A} - u_{y}^{C} -l_{y}u_{x}^{RP2}=0$}   \\
\hline
\end{tabular}
\label{corner}
\end{table}

\begin{table}[ht]
\caption{Constraint equation on edges}
\centering
\begin{tabular}{c@{\hskip 0.5in}c} 

\hline
\textbf{BD and AC} & \textbf{AB and CD} \\ 
\hline
$u_{x}^{BD} - u_{x}^{AC} -l_{x}u_{x}^{RP1}=0$& $u_{y}^{AB} - u_{y}^{CD} -l_{y}u_{x}^{RP2}=0$\\

$u_{y}^{BD} - u_{y}^{AC} -\frac{l_{x}}{2}u_{x}^{RP3}=0$&$u_{x}^{AB} - u_{x}^{CD} -\frac{l_{x}}{2}u_{x}^{RP3}=0$     \\
\hline
\end{tabular}
\label{edge}
\end{table}

\begin{table}[ht]
\caption{Boundary condition for determining elastoplastic behavior}
\centering
\begin{tabular}{c@{\hskip 0.2in}c} 
\hline
\textbf{Applied strain} & \textbf{Boundary condition} \\ 
\hline
Uniform  & $u_{x}^{RP1} = \epsilon_{11}$ \\

Compressive  & $u_{x}^{RP2}=0$ \\

stain in x-direction & $u_{x}^{RP3}=0$ \\
\hline
\end{tabular}
\label{BC}
\end{table}

\begin{figure}[htbp]
\centering
\includegraphics[width=\columnwidth]{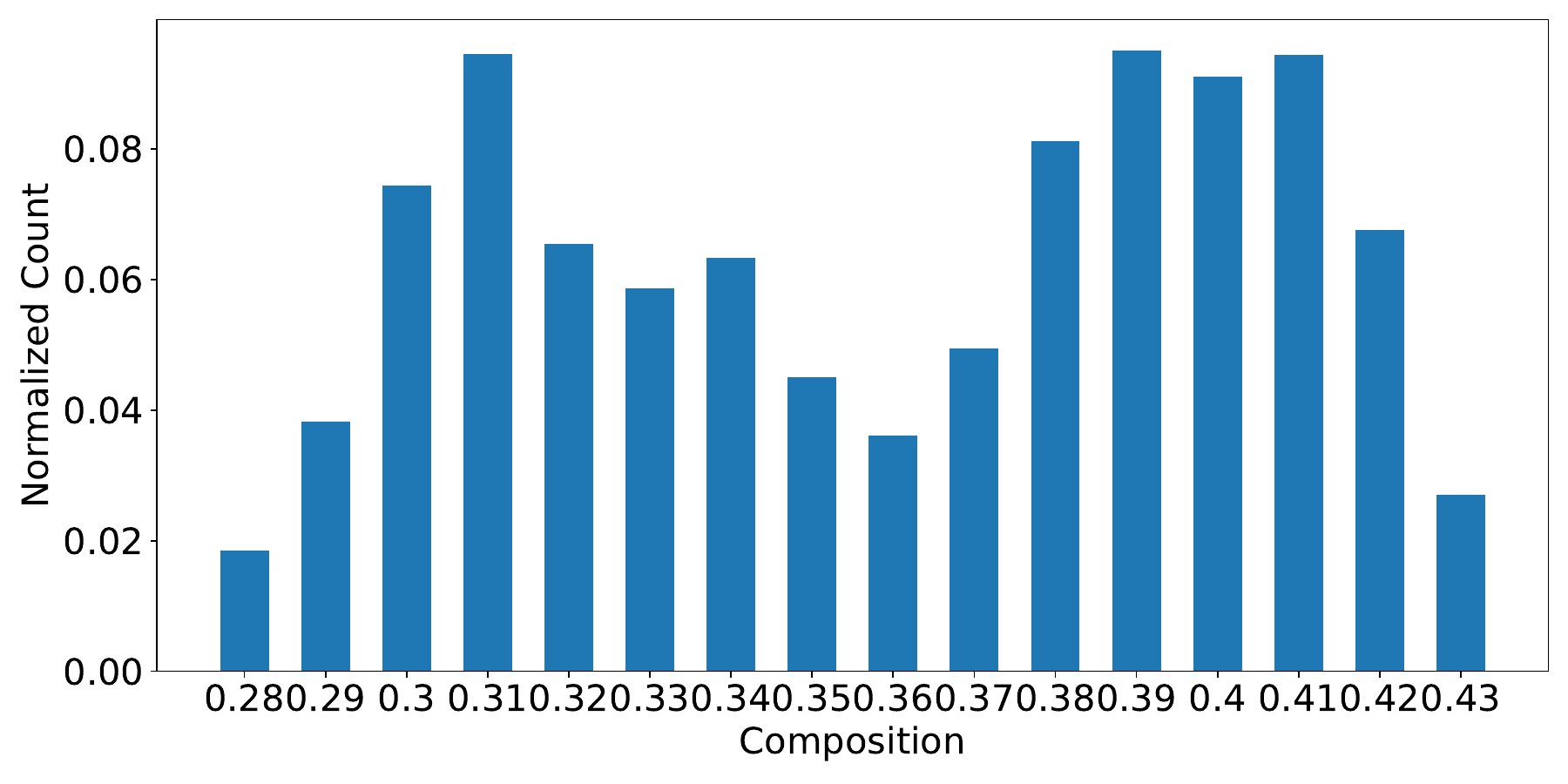}
\caption{Frequency distribution of different compositions among the GAN-predicted images.}
\label{compdist}
\end{figure}

\begin{figure*}[ht]
\centering
\includegraphics[width=0.9\textwidth]{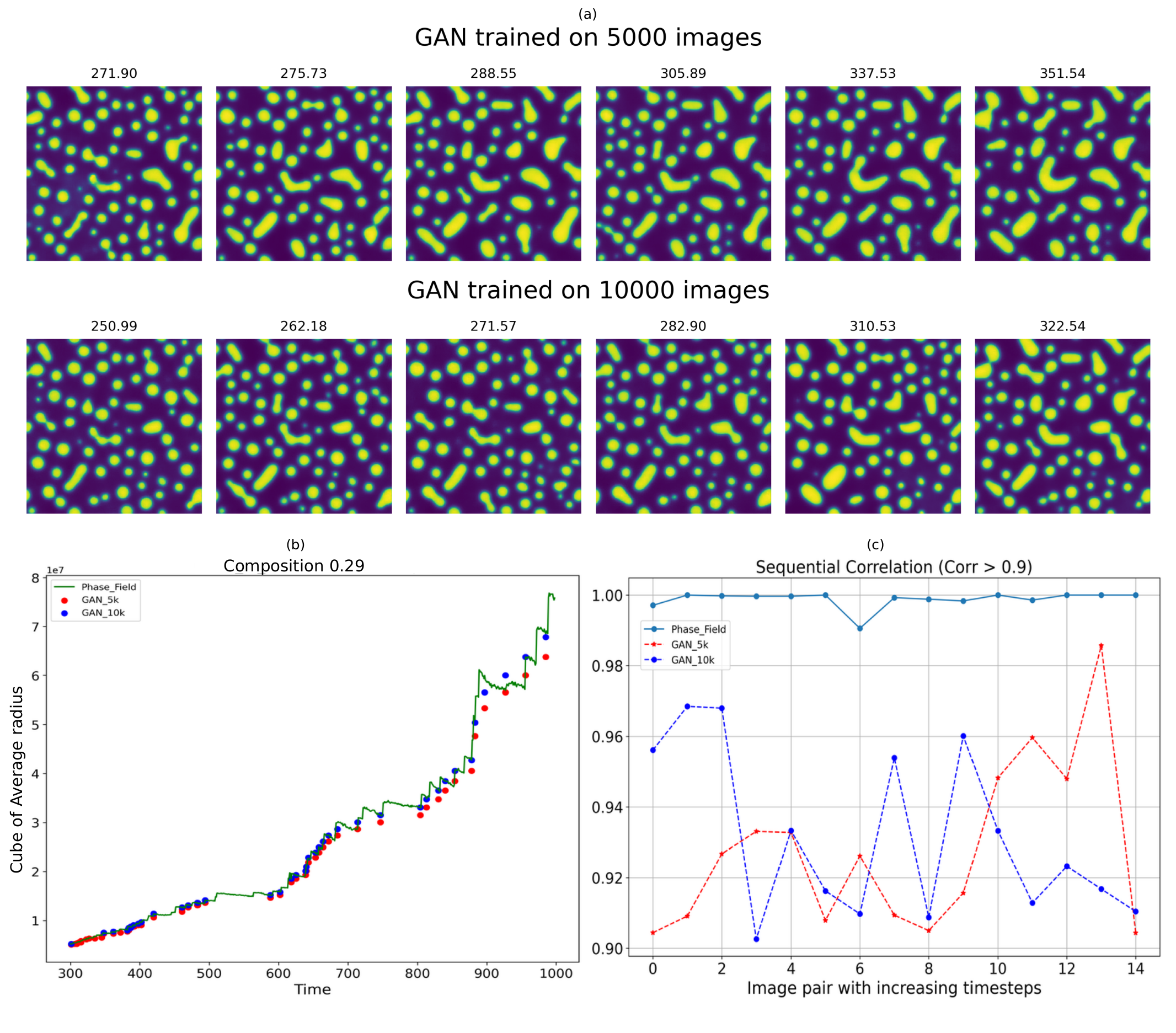}
\caption{(a) Synthetic microstructures generated with GAN, using noise as input for initial composition $\phi_A=0.29$. The top and bottom row shows the output of GAN trained with 5000 and 10000 images, respectively. The average precipitate radius is given in angstrom with every image. (b) Cube of average precipitate radius plotted as a function of time: synthetic microstructures showing the same time dependence as the phase-field microstructures. (c) Synthetic microstructures showing high sequential correlation.}
\label{fig3}
\end{figure*}

Abaqus solver output variation of stress and strain for a given meshed microstructure at each time step, as shown in Figure~\ref{methodology} (e) and (f), respectively. The distribution of true normal stress ($\sigma_{xx}$) and true normal strain ($\epsilon_{xx}$)in the x-direction is shown in the last time step. The area-averaged stress components and strain components are calculated from a given stress-strain distribution at each time step during post-processing from Eq.~\ref{avg_stress} and Eq.~\ref{avg_strain} to compute the true stress-strain curve for deformed microstructure. 
\begin{equation}
\overline{\sigma_{ij}}=\frac{1}{A}\int\sigma_{ij}dA = \frac{1}{A}\sum_{n=1}^{nel}\sigma_{ij}^{n}A^{n}
\label{avg_stress}
\end{equation}
\begin{equation}
\overline{\epsilon_{ij}}=\frac{1}{A}\int\epsilon_{ij}dA = \frac{1}{A}\sum_{n=1}^{nel}\epsilon_{ij}^{n}A^{n}
\label{avg_strain}
\end{equation}
In the above equations, $A$ is the area of the microstructure; $\overline{\sigma_{ij}}, \overline{\epsilon_{ij}}$ are the area-averaged values of stress and strain, respectively; $nel$ is the total number of finite elements of the meshed microstructure; $\sigma_{ij}^{n}$ is $ij^{th}$ component of the stress tensor calculated in the $n^{th}$ element, and $\epsilon_{ij}^{n}$ is $ij^{th}$ component of the strain tensor calculated in the $n^{th}$ element.

\begin{figure*}[ht]
\centering
\includegraphics[width=0.9\textwidth]{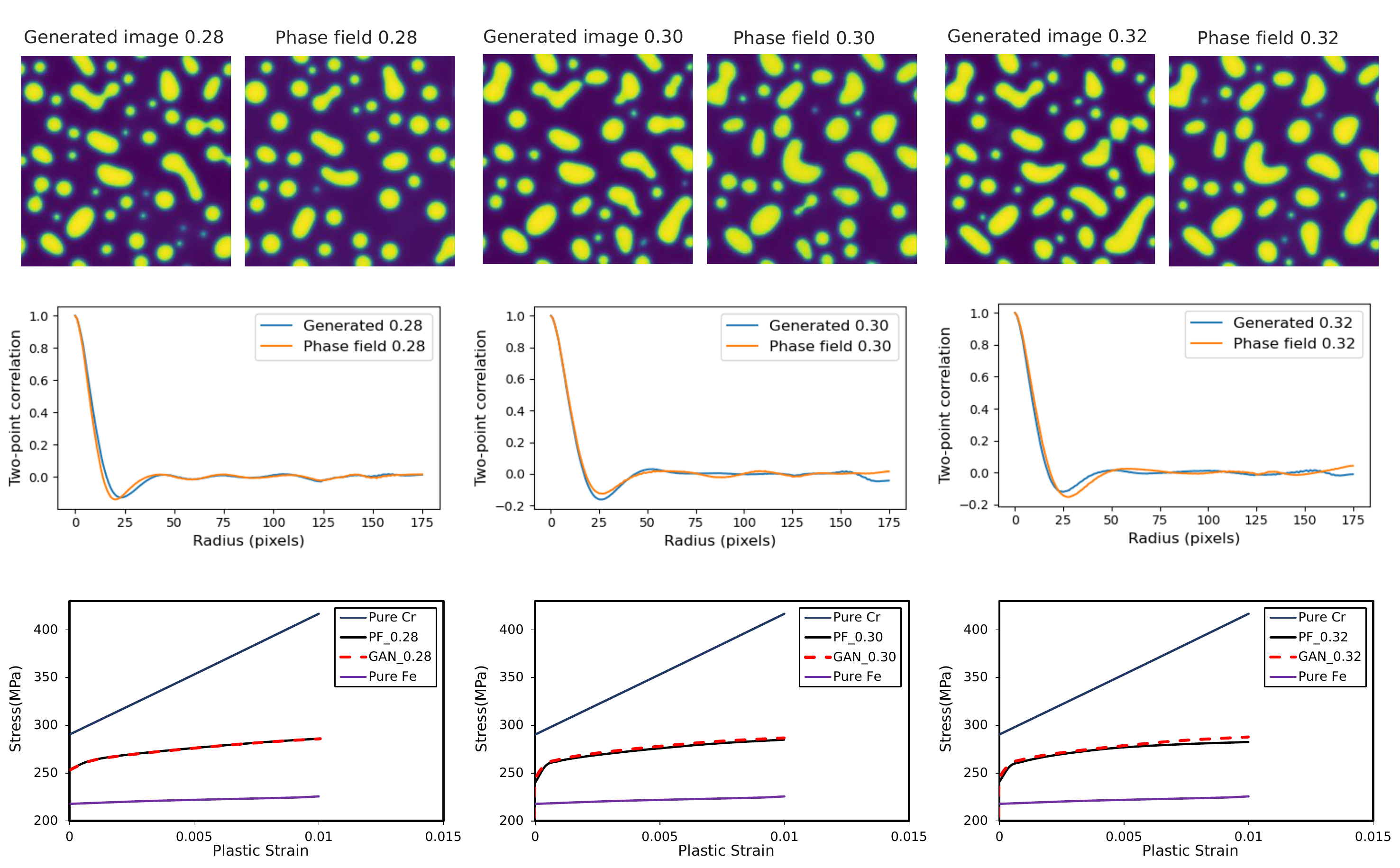}
\caption{Top row: Phase-field and GAN-predicted microstructures having different compositions and similar particle size distribution. Middle row: Comparison of two-point correlation. Last row: True stress-plastic strain curves, calculated using the phase-field and GAN-generated microstructures. Curves for pure iron and chromium are also given as a reference.}
\label{Fig-4}
\end{figure*}

\section{Results and Discussions}
As mentioned previously, we generate 20000 microstructure images, spanning a wide range of compositions and time steps, using a phase-field model. We shuffle these 20000 microstructures and randomly divide them into four sets, each containing 5000 images. The four sets have standard deviations as follows [0.3136, 0.3140, 0.3138, 0.3155], and we chose the fourth set (which has the maximum standard deviation) for training GAN. Since the outcome of GAN may depend on the number of images used for training, we also construct a larger dataset of 10,000 training images, again selected randomly from the set of 20000 phase-field microstructures. After training the model, it takes less than ten seconds to generate 5000 synthetic microstructures. It is desirable to have a fair representation of every composition among the microstructures predicted by GAN. We plot the frequency of each composition based on 5000 images generated by GAN [Figure~\ref{compdist}]. The plot clearly shows the diversity of the synthetic microstructures in terms of composition.

Since we generate the microstructures from a state of pure noise, additional efforts are needed to group them in terms of composition and sort them chronologically. First, we find the composition of the generated microstructures by calculating the average pixel value for a given microstructure. Next, we methodically sort the microstructures in ascending order for a given composition based on the average particle size. We illustrate the microstructures generated by a model that has been trained with a dataset comprising 5,000 and 10,000 images [see Figure~\ref{fig3}(a), the first and second row] for a composition $\phi_A=0.29$. Interestingly, GAN is capable of generating realistic microstructures even with a smaller dataset of 5,000 training images.

During GAN training for microstructure generation, standard hyperparameter configurations, such as a learning rate (LR) of \(2 \times 10^{-4}\) with the Adam optimizer and a batch size of 32 were employed and found sufficient for achieving stable convergence and high-quality outputs. Given the model’s consistent training behavior and ability to generate visually and structurally realistic microstructures under these default settings, exhaustive hyperparameter optimization was not deemed necessary.

While generating a set of visually realistic microstructures is an encouraging sign, one needs to probe further whether GAN can accurately capture the underlying physics of microstructure evolution. Let us first verify whether the synthetic microstructures generated by GAN follow the theoretically predicted $r\propto t^{1/3}$ coarsening rate, where $r$ is the average radius, and $t$ is time. Figure~\ref{fig3}(b) depicts the cube of the average radius as a function of time. The green line is obtained by analyzing the phase field microstructures. The red and blue dots are obtained by analyzing the synthetic microstructures generated by GAN, trained with 5,000 and 10,000 images. Clearly, GAN is capable of producing synthetic microstructures, which obey the theoretically predicted coarsening rate. A similar analysis is presented for other compositions in Figures~S2, S3, and S4 of the Supporting Information.

Encouraged by the success of GAN in generating realistic microstructures, we further analyze the sequential correlation between two consecutive pairs of images (like first-second, second-third, third-fourth, etc.). Sequential correlation is a statistical measure used to evaluate the degree of similarity between sequences, specifically focusing on consecutive items within these sequences. In the context of analyzing microstructures, this metric assesses how closely related consecutive pairs of images are in terms of their structural features, which is critical when studying processes that evolve over time, such as microstructure formation and transformation. Mathematically, the sequential correlation between two images \( I_{m} \) and \( I_{m+n} \) can be expressed using the Pearson correlation coefficient formula:
\begin{equation}
\label{seqcor}
r = \frac{\sum (I_{m} - \overline{I}_{m})(I_{m+n} - \overline{I}_{m+n})}{\sqrt{\sum (I_{m} - \overline{I}_{m})^2 \sum (I_{m+n} - \overline{I}_{m+n})^2}},    
\end{equation}
where \( I_{m} \) and \( I_{m+n} \) are the pixel values of $m^{th}$ and $(m+n)^{th}$ images, \( \overline{I}_{m} \) and \( \overline{I}_{m+n} \) are the mean pixel values of $m^{th}$ and $(m+n)^{th}$ images. The sums are computed over all pixels in the images. For two consecutive pairs of images, $n=1$.

A high sequential correlation coefficient, close to 1, indicates that the microstructures in consecutive images are similar, suggesting that the transitions between these states are continuous. This measure is particularly useful in the context of GAN-generated data to ensure that the synthetic sequences maintain temporal consistency, reflecting the natural progression of microstructure evolution. As shown in Figure~\ref{fig3}(c), phase field microstructures have a very high sequential correlation, close to 1, indicating that they originate from the same nucleation point. Interestingly, synthetic microstructures generated by GAN also have reasonably high sequential correlation ($>0.9$). This finding is pivotal as it confirms the model’s reliability in producing sequentially dependent microstructures, which is essential for temporal evolution studies.

As mentioned previously, one of the motivations for this work is to generate realistic synthetic microstructures for FEA-based property calculations. Thus, to assess the fidelity of our model, we use the GAN-generated microstructures for property calculations and compare them with the property values obtained from phase-field microstructures. We perform virtual mechanical testing in uniform compression using three compositions, 0.28, 0.30, and 0.32, using microstructures from GAN and phase-field calculations. The tested microstructures from GAN and the phase-field method contain a similar number of particles. 

Before comparing their properties, we first visually and statistically compare the microstructures [Figure~\ref{Fig-4}]. The first row of the figure displays side-by-side comparisons of the microstructures. The quality of generated microstructures was quantitatively assessed using the Fréchet Inception Distance (FID), a widely used metric for evaluating generative models. We computed FID scores between GAN-generated and phase-field microstructures across different compositions: 0.28 (FID: 73.84), 0.30 (FID: 59.75), and 0.32 (FID: 73.88). These moderate values indicate that our model has learned the underlying distribution rather than simply memorizing the training examples, confirming its ability to generate novel microstructures with appropriate statistical variations while maintaining physical consistency.

The second row of Figure~\ref{Fig-4} presents the comparative analysis of the two-point correlation between GAN-generated and phase-field microstructures with similar composition and average particle size, revealing that both GAN and phase-field images exhibit comparable spatial relationships. The two-point correlation function \( S_2(r) \) for a given microstructure is defined as the probability that two points separated by a distance \( r \) both belong to the same phase. Mathematically, the function can be expressed as
\begin{equation}
S_2(r) = \frac{1}{N(r)} \sum_{i=1}^{N(r)} \delta(x_i) \delta(x_i + r),    
\end{equation}
where \( \delta(x) \) is an indicator function that is 1 if \( x \) belongs to the phase of interest and 0 otherwise; \( x_i \) is the position of the $i^{th}$ point; \( x_i + r \) is a point at distance \( r \) from \( x_i \). The sum is over all pairs of points separated by distance \( r \), and \( N(r) \) is the number of such pairs. The above formula calculates the normalized sum of products of the indicator function over all pairs of points at a given separation. This helps determine how well the GAN can replicate the spatial distribution of phases observed in physics-based models like phase-field simulations. A close match in the two-point correlation functions between GAN and phase-field microstructures [Figure~\ref{Fig-4}] indicates that the GAN can accurately capture the spatial characteristics, which is essential to the material's properties. The results for virtual mechanical testing are shown in the last row of Figure~\ref{Fig-4}. As mentioned previously, the stress-strain curves obtained from the phase-field and GAN microstructure are within the limits of the rule of mixture for composites. The figure depicts an excellent agreement between the stress-strain curves obtained from the GAN  and phase-field microstructures. A more detailed error analysis is presented in Section~S2-D and Figure~S3 of the Supporting Information.

\section{Discussion and Limitations}
\label{Discussion and Limitations}

While our GAN-based approach demonstrates significant advantages in computational efficiency and high-quality microstructure generation, several limitations warrant discussion. First, the model's capabilities are inherently bounded by the diversity and representativeness of the training dataset. Our current implementation relies on phase-field simulations with simplified thermodynamic parameters, which may not capture the full complexity of real-world microstructural evolution under varying processing conditions. Second, the present study focuses exclusively on binary systems with spinodal decomposition. Extending to more complex multi-component systems or different microstructural evolution mechanisms (e.g., dendritic solidification or precipitation hardening) would require additional architectural modifications and significantly larger training datasets.

Different GAN architectures offer various advantages for microstructure generation, each with inherent limitations - vanilla GANs often suffer from mode collapse and training instability when dealing with complex microstructural features, conditional GANs require explicit labeled data which can be challenging to obtain for continuous microstructural evolution sequences, and cycleGANs introduce computational overhead that may not be justified for single-domain generation tasks. Our selection of Wasserstein GAN was motivated by its superior training stability, improved gradient properties, and ability to capture complex distributions without the drawbacks above, making it particularly well-suited for capturing the subtle details of microstructural evolution in materials science applications.

From a computational perspective, although GANs offer remarkable efficiency gains compared to phase-field simulations, they still require substantial computational resources during the training phase, particularly when scaling to 3D microstructures. Furthermore, while our current approach demonstrates accurate reproduction of microstructural features and mechanical properties, the GAN model lacks explicit physical constraints, potentially generating microstructures that, while statistically realistic, might violate certain physical laws. More research is needed in this area.

Additionally, the model's ability to generate microstructures outside the composition and time ranges present in the training data remains limited, constraining its extrapolative capabilities for materials discovery. Despite these constraints, this approach proves particularly valuable for generating microstructures with minimal software and coding expertise. The ease of implementation, once trained, allows materials scientists with limited computational backgrounds to rapidly generate diverse microstructures for various studies. This accessibility, combined with the demonstrated computational efficiency and property prediction accuracy, suggests that GAN-based approaches offer a promising framework for accelerating microstructure-based materials design, particularly when coupled with physics-informed constraints and more diverse training datasets.

\section{Conclusions}
\label{conclusion}
We demonstrate the capability of GAN, one of the most popular generative models, to capture the complexity of spatiotemporal evolution in the context of microstructure modeling. Using microstructures obtained from phase-field simulation, we train a GAN model, which can generate high-quality synthetic microstructures. The generated microstructure dataset is diverse in terms of compositional (varying over a wide range) as well as temporal (containing initial as well as coarsened microstructures) variation. We also calculate the mechanical properties using the finite element method on synthetic and actual phase field microstructures and find a good match. 

In the era of integrated computational materials engineering, such models can significantly accelerate the development of new materials. Using the GAN model, one can generate a large number of realistic microstructures in a short time. For example, in our case, generating 20000 realistic synthetic images takes 40 seconds in an Intel\textregistered{} Xeon\textregistered{} Gold 6338N CPU (2.20~GHz), having 128 GB RAM. In comparison, it takes 1800 seconds for the phase-field model (with OpenMP parallelization~\cite{Linda_2024}) to generate microstructures of the same quality and quantity using a machine of similar hardware configuration. The capability of generating such a large number of microstructures at a minimal computational cost is advantageous for further property predictions, for example, using FEM calculations. One can leverage such a database of diverse synthetic microstructures for statistically reliable predictions.

Looking ahead, we envision several promising expansion paths for this method. The GAN framework demonstrated here could be extended to multiscale material modeling by incorporating hierarchical architectures that capture features at different length scales simultaneously. Additionally, reinforcement learning techniques could be integrated with our generative approach to enable controlled microstructure synthesis with targeted properties, effectively steering the generative process toward desired mechanical, thermal, or electrical characteristics. Furthermore, physically driven generative modeling, which embeds physical laws and constraints directly into the learning process, represents another compelling direction. Such physics-informed GANs would ensure not only visual fidelity but also physical consistency of the generated microstructures. These advancements would bridge the gap between data-driven approaches and physics-based modeling, potentially revolutionizing computational materials design by combining the efficiency of machine learning with the reliability of physical principles.

\section{Data availability}
Code is available in the author's \href{https://github.com/owaisahmad18/Synthetic-microstructure-generation-with-GAN/blob/main/Sythetic_GAN.ipynb}{github} page. Training data is available in \href{https://zenodo.org/records/13823146}{this link} and the trained model is available in \href{https://zenodo.org/records/13811336}{this link}.

\section{Acknowledgements}
Authors acknowledge Center for Development of Advanced Computing (C-DAC) for funding (Project No. Meity/R\&D/HPC/2(1)/2014) and National Supercomputing Mission (NSM) for providing computing resources of Param Sanganak at IIT Kanpur, which is implemented by C-DAC and supported by the Ministry of Electronics and Information Technology (MeitY) and Department of Science and Technology (DST), Government of India. The authors also thank ICME National Hub and CC, IIT Kanpur, for providing the HPC facility.


\begin{thebibliography}{10}

\bibitem{ROTERS20101152}
F.~Roters, P.~Eisenlohr, L.~Hantcherli, D.D. Tjahjanto, T.R. Bieler, and
  D.~Raabe.
\newblock Overview of constitutive laws, kinematics, homogenization and
  multiscale methods in crystal plasticity finite-element modeling: Theory,
  experiments, applications.
\newblock {\em Acta Materialia}, 58(4):1152--1211, 2010.

\bibitem{abaqus}
Michael Smith.
\newblock {\em ABAQUS/Standard User's Manual, Version 6.9}.
\newblock Dassault Syst{\'e}mes Simulia Corp, United States, 2009.

\bibitem{Langer200115}
S.A. Langer, E.R. Fuller, and W.C. Carter.
\newblock Oof: an image-based finite-element analysis of material
  microstructures.
\newblock {\em Computing in Science \& Engineering}, 3(3):15--23, 2001.

\bibitem{FROMM20125984}
Bradley~S. Fromm, Kunok Chang, David~L. McDowell, Long-Qing Chen, and Hamid
  Garmestani.
\newblock Linking phase-field and finite-element modeling for
  process–structure–property relations of a ni-base superalloy.
\newblock {\em Acta Materialia}, 60(17):5984--5999, 2012.

\bibitem{Linda_2024}
Albert Linda, Ankit~Singh Negi, Vishal Panwar, Rupesh Chafle, Somnath Bhowmick,
  Kaushik Das, and Rajdip Mukherjee.
\newblock $\mu$2mech: A software package combining microstructure modeling and
  mechanical property prediction.
\newblock {\em Physica Scripta}, 99(5):055256, apr 2024.

\bibitem{LIU2020102614}
Wenqi Liu, Junhe Lian, Nikolaos Aravas, and Sebastian Münstermann.
\newblock A strategy for synthetic microstructure generation and crystal
  plasticity parameter calibration of fine-grain-structured dual-phase steel.
\newblock {\em International Journal of Plasticity}, 126:102614, 2020.

\bibitem{dream}
Michael~A. Groeber and Michael~A. Jackson.
\newblock Dream.3d: A digital representation environment for the analysis of
  microstructure in 3d.
\newblock {\em Integrating Materials and Manufacturing Innovation}, 3:56--72,
  2014.

\bibitem{mukherjee2009}
R.~Mukherjee, T.A. Abinandanan, and M.P. Gururajan.
\newblock Phase field study of precipitate growth: Effect of misfit strain and
  interface curvature.
\newblock {\em Acta Materialia}, 57(13):3947--3954, 2009.

\bibitem{Mukhrjee2010}
R.~Mukherjee, T.A. Abinandanan, and M.P. Gururajan.
\newblock Precipitate growth with composition-dependent diffusivity: Comparison
  between theory and phase field simulations.
\newblock {\em Scripta Materialia}, 62(2):85--88, 2010.

\bibitem{Gururajan2007}
M.P. Gururajan and T.A. Abinandanan.
\newblock Phase field study of precipitate rafting under a uniaxial stress.
\newblock {\em Acta Materialia}, 55(15):5015--5026, 2007.

\bibitem{CHANG201767}
Kunok Chang, Long-Qing Chen, Carl~E Krill, and Nele Moelans.
\newblock {Effect of strong nonuniformity in grain boundary energy on 3-D grain
  growth behavior: A phase-field simulation study}.
\newblock {\em Computational Materials Science}, 127:67--77, 2017.

\bibitem{verma_mukherjee}
M.~Verma and R.~Mukherjee.
\newblock Grain growth stagnation in solid state thin films: A phase-field
  study.
\newblock {\em Journal of Applied Physics}, 130(2):025305, 2021.

\bibitem{PhysRevLett.86.842}
C.~E. Krill, L.~Helfen, D.~Michels, H.~Natter, A.~Fitch, O.~Masson, and
  R.~Birringer.
\newblock Size-dependent grain-growth kinetics observed in nanocrystalline fe.
\newblock {\em Phys. Rev. Lett.}, 86:842--845, Jan 2001.

\bibitem{MOLNAR20126961}
David Molnar, Rajdip Mukherjee, Abhik Choudhury, Alejandro Mora, Peter Binkele,
  Michael Selzer, Britta Nestler, and Siegfried Schmauder.
\newblock Multiscale simulations on the coarsening of cu-rich precipitates in
  $\alpha$-fe using kinetic monte carlo, molecular dynamics and phase-field
  simulations.
\newblock {\em Acta Materialia}, 60(20):6961--6971, 2012.

\bibitem{ZHAO20191044}
Yuhong Zhao, Bing Zhang, Hua Hou, Weipeng Chen, and Meng Wang.
\newblock {Phase-field simulation for the evolution of solid/liquid interface
  front in directional solidification process}.
\newblock {\em Journal of Materials Science \& Technology}, 35(6):1044--1052,
  2019.

\bibitem{chatterjee2008phase}
Subhradeep Chatterjee, TA~Abinandanan, and Kamanio Chattopadhyay.
\newblock Phase-field simulation of fusion interface events during
  solidification of dissimilar welds: effect of composition inhomogeneity.
\newblock {\em Metallurgical and Materials Transactions A}, 39:1638--1646,
  2008.

\bibitem{CHAFLE}
Rupesh Chafle, Somnath Bhowmick, and Rajdip Mukherjee.
\newblock Effect of co-existing external fields on a binary spinodal system: A
  phase-field study.
\newblock {\em Journal of Physics and Chemistry of Solids}, 132:236--243, 2019.

\bibitem{RAGHAVAN2021}
Rahul Raghavan, William Farmer, Leslie~T. Mushongera, and Kumar Ankit.
\newblock Multiphysics approaches for modeling nanostructural evolution during
  physical vapor deposition of phase-separating alloy films.
\newblock {\em Computational Materials Science}, 199:110724, 2021.

\bibitem{vondrous2014parallel}
Alexander Vondrous, Michael Selzer, Johannes H{\"o}tzer, and Britta Nestler.
\newblock Parallel computing for phase-field models.
\newblock {\em The International journal of high performance computing
  applications}, 28(1):61--72, 2014.

\bibitem{wang2020}
Anthony Yu-Tung Wang, Ryan~J. Murdock, Steven~K. Kauwe, Anton~O. Oliynyk,
  Aleksander Gurlo, Jakoah Brgoch, Kristin~A. Persson, and Taylor~D. Sparks.
\newblock Machine learning for materials scientists: An introductory guide
  toward best practices.
\newblock {\em Chemistry of Materials}, 32(12):4954--4965, 2020.

\bibitem{xue2022physics}
Tianju Xue, Zhengtao Gan, Shuheng Liao, and Jian Cao.
\newblock Physics-embedded graph network for accelerating phase-field
  simulation of microstructure evolution in additive manufacturing.
\newblock {\em npj Computational Materials}, 8(1):201, 2022.

\bibitem{oommen2022learning}
Vivek Oommen, Khemraj Shukla, Somdatta Goswami, R{\'e}mi Dingreville, and
  George~Em Karniadakis.
\newblock Learning two-phase microstructure evolution using neural operators
  and autoencoder architectures.
\newblock {\em npj Computational Materials}, 8(1):190, 2022.

\bibitem{WU2023}
Peichen Wu, Ashif~Sikandar Iquebal, and Kumar Ankit.
\newblock Emulating microstructural evolution during spinodal decomposition
  using a tensor decomposed convolutional and recurrent neural network.
\newblock {\em Computational Materials Science}, 224:112187, 2023.

\bibitem{hu2022accelerating}
C~Hu, S~Martin, and R~Dingreville.
\newblock Accelerating phase-field predictions via recurrent neural networks
  learning the microstructure evolution in latent space.
\newblock {\em Computer Methods in Applied Mechanics and Engineering},
  397:115128, 2022.

\bibitem{montes2021accelerating}
David Montes~de Oca~Zapiain, James~A Stewart, and R{\'e}mi Dingreville.
\newblock Accelerating phase-field-based microstructure evolution predictions
  via surrogate models trained by machine learning methods.
\newblock {\em npj Computational Materials}, 7(1):1--11, 2021.

\bibitem{TIWARI2025113518}
Saurabh Tiwari, Prathamesh Satpute, and Supriyo Ghosh.
\newblock Time series forecasting of multiphase microstructure evolution using
  deep learning.
\newblock {\em Computational Materials Science}, 247:113518, 2025.

\bibitem{ahmad2023accelerating}
Owais Ahmad, Naveen Kumar, Rajdip Mukherjee, and Somnath Bhowmick.
\newblock Accelerating microstructure modeling via machine learning: A method
  combining autoencoder and convlstm.
\newblock {\em Phys. Rev. Mater.}, 7:083802, Aug 2023.

\bibitem{ahmad2024}
Owais Ahmad, Rakesh Maurya, Rajdip Mukherjee, and Somnath Bhowmick.
\newblock Integrated phase field and machine learning study of microstructure
  evolution during interface-controlled spinodal decomposition.
\newblock {\em Solid State Phenomena}, 357:101--106, 6 2024.

\bibitem{westphal2024generative}
Erik Westphal and Hermann Seitz.
\newblock Generative artificial intelligence: Analyzing its future applications
  in additive manufacturing.
\newblock {\em Big Data and Cognitive Computing}, 8(7), 2024.

\bibitem{LIU2023798}
Yue Liu, Zhengwei Yang, Zhenyao Yu, Zitu Liu, Dahui Liu, Hailong Lin, Mingqing
  Li, Shuchang Ma, Maxim Avdeev, and Siqi Shi.
\newblock Generative artificial intelligence and its applications in materials
  science: Current situation and future perspectives.
\newblock {\em Journal of Materiomics}, 9(4):798--816, 2023.

\bibitem{Vinodhini_2025}
M~Vinodhini and Sujatha Rajkumar.
\newblock A deep learning approach for predicting and optimizing the v2x
  network parameters for sustainable smart transportation systems.
\newblock {\em Engineering Research Express}, 7(1):015268, feb 2025.

\bibitem{doi:10.1063/1.1744102}
John~W Cahn and John~E Hilliard.
\newblock {Free Energy of a Nonuniform System. I. Interfacial Free Energy}.
\newblock {\em The Journal of Chemical Physics}, 28(2):258--267, 1958.

\bibitem{cahn1961spinodal}
John~W Cahn.
\newblock On spinodal decomposition.
\newblock {\em Acta metallurgica}, 9(9):795--801, 1961.

\bibitem{narikawa2022generative}
Ryuichi Narikawa, Yoshihito Fukatsu, Zhi-Lei Wang, Toshio Ogawa, Yoshitaka
  Adachi, Yuji Tanaka, and Shin Ishikawa.
\newblock Generative adversarial networks-based synthetic microstructures for
  data-driven materials design.
\newblock {\em Advanced Theory and Simulations}, 5(5):2100470, 2022.

\bibitem{HENKES2022115497}
Alexander Henkes and Henning Wessels.
\newblock Three-dimensional microstructure generation using generative
  adversarial neural networks in the context of continuum micromechanics.
\newblock {\em Computer Methods in Applied Mechanics and Engineering},
  400:115497, 2022.

\bibitem{azqadan2023predictive}
Erfan Azqadan, Hamid Jahed, and Arash Arami.
\newblock Predictive microstructure image generation using denoising diffusion
  probabilistic models.
\newblock {\em Acta Materialia}, 261:119406, 2023.

\bibitem{Fritz2017176}
R.~Fritz, D.~Wimler, A.~Leitner, V.~Maier-Kiener, and D.~Kiener.
\newblock Dominating deformation mechanisms in ultrafine-grained chromium
  across length scales and temperatures.
\newblock {\em Acta Materialia}, 140:176 – 187, 2017.

\bibitem{Yabe20141342}
Shintaro Yabe, Motoki Terano, and Masahiko Yoshino.
\newblock Plane strain compression test and simple shear test of single crystal
  pure iron.
\newblock {\em Procedia Engineering}, 81:1342--1347, 2014.
\newblock 11th International Conference on Technology of Plasticity, ICTP 2014,
  19-24 October 2014, Nagoya Congress Center, Nagoya, Japan.

\bibitem{hertzberg2012deformation}
R.W. Hertzberg, R.P. Vinci, and J.L. Hertzberg.
\newblock {\em Deformation and Fracture Mechanics of Engineering Materials}.
\newblock Wiley, 2012.

\end{thebibliography}

\end{document}